# Perspectives on Einstein's scientific works in Milan

Christian Bracco

*Syrte, Observatoire de Paris, PSL Research University, CNRS, Sorbonne Universités, UPMC Univ. Paris 06, LNE, 61 avenue de l'Observatoire,*
*Paris, 75014, France*
*E-mail: Christian.Bracco@obspm.fr*

The Milanese period in Albert Einstein's life is a key one for the understanding of the development of his scientific questioning. While being a student in Zürich from 1896, Einstein returned regularly to Milan to meet his family for the holidays. There, he could work on the most recent articles in physics at the rich library of the Lombardo Institute, Academy of Sciences and Letters. Hopefully, this new perspective will help to trace back more accurately a few of Einstein's scientific ideas, such as the need to expand his first doctoral work on molecular forces to weakly compressed gases or as to conceive a first idea of light quanta.

## 1. Introduction

Albert Einstein (1879-1955) spent about three months a year in Milan between 1896, the year he entered the ETH in Zürich, and 1901, during his PhD studies. These family trips match the university vacation periods, and letters addressed to his friend Mileva Marić were sent in September, October, March and April. They constitute the main testimony to understand his scientific questionings. They were edited in the first volume of the *Collected Papers of Albert Einstein* in 1987 [1] then were separately published in 1993 [2]. Regarding the nature of light, the analysis conducted on the basis of these letters lead Jürgen Renn to think that, as soon as 1901, Einstein could have had a first idea - not yet formalized - of light quanta [3]. In 1900, Mileva expected Einstein to defend his thesis in relation to molecular forces on Easter 1901 [1, Doc. 85]. However this was the time when his scientific questionings became pressing. The letters to Mileva, six in total, written in Milan, from March 23rd to April 30th, are, from this point of view, the most important ones, with a letter to Grossmann on April 14th, and a letter to Mileva written in Winterthur in May. The new elements that I present ensue from Einstein's scientific environment in Milan [4] and its possible scientific consequences [5]. To go beyond the knowledge of the incomplete elements included in his letters, it is important to determine the place where Einstein used to work, which will allow us to learn about the biographical sources at his disposal, when he continued in Milan the work he started in Zürich. The presence at his side of his friend Michele Besso between 1899 and 1901 should also be noted. In March 1901, their regular scientific discussions focused on topics such as « *the fundamental separation of luminiferous ether and matter, the* *definition of absolute rest, molecular forces, surface phenomena, dissociation"* [1, Doc. 96]. It is worth noting that the letter addressed to Einstein by Besso of February 1903 [6], the older one published, shows that their informal collaboration was based on an exhaustive scientific bibliography of the topics approached. The appreciation of the library where they would work in Milan thus becomes a key element.

## 2.  Scientific perspectives

### 2.1.  *The library of the Lombardo Institute, Academia of Science and Letters*

On April 4[th] 1901, Einstein studied Drude's electron theory [1, Doc. 96] "*I have in my hands a study by Paul Drude on the electron theory*" and he concluded "*But I must be off to the library, otherwise it will be getting to late*". He had obviously read Drude's work in this library, since he wrote the next week: "*Last week I studied* […] *the electron theory of metals in the library*" [1, Doc. 97]. Yet, the only library at the time in Milan that owns the *Annalen der Physik* journal, in which Drude's work is published [1, Doc. 96, n. 4 ], was at the Lombardo Institute, Academia of Science and Letters. It is situated in the Brera palace, five hundred meters away from the Einstein residence, at 21 via Bigli. Several other correlations between the letters and the library collection can be highlighted, and besides, an extract from Rudolph Kayser's biography confirms Einstein's presence at the Brera Palace [4]. Einstein and Besso, both starting a thesis, could not have missed the reading of a nearly seven hundred pages edition that the *Archives néerlandaises* journal devoted to the *Festschrift* for Lorentz of December 11[th] 1900, with articles written by Planck, Boltzmann, Wien, Wiechert, Kaufmann, Van der Waals Jr, Voigt, Zeeman, etc. This document has been owned by the library since January 31[st] 1901.

### 2.2.  *The reorientation of Einstein's doctoral dissertation in April 1901: Reinganum's paper in Lorentz' Festschrift*

On April 14[th], Einstein wrote to his friend Marcel Grossmann: "*As for science, I have a few splendid ideas, which only now need proper incubation. I am now convinced that my theory of atomic attraction forces can also be extended to gases* […]. *That will also bring the problem of the inner kinship between molecular forces and Newtonian action-at-a-distance forces much nearer to its solution* […]. *I shall utilize the already existing results in my doctoral dissertation*"[1, Doc. 100]. Using analogous terms, he wrote to Mileva the next day, mentioning "*an extremely lucky idea, which will make it possible to apply our theory of molecular forces to gases as well*", that would lead to  "*know almost as much about molecular forces as about gravitational forces, and only the law of radius will still remain unknown*" [1, Doc. 101]. He came back to this topic on April 30[th] and added a condition: "*if the mathematically so unclear concept of molecular size does not show up as well in the formation of trajectories of molecules coming close to each other as well, but the molecule can be conceived as center of force*" [1, Doc.]. Max Reinganum's article entitled *On molecular forces in weakly compressed gases* [7] is surely related to Einstein's enthusiasm, who found in this work a point of view matching his own and opposed to Boltzmann's, that is to say that the molecules size does not play any role. Reinganum considered the molecules in the gas as force centers subjected to a "planetary" interaction in $r^{-4}$. It should be noted that this idea is a consistent continuation of Einstein's work on capillarity that had just been published in the *Annalen*. He submitted his first PhD manuscript to Alfred Keiner on this basis, in December 1901, as Mileva indicated: *Albert has written a magnificent study, which he submitted as his dissertation* […] *It deals with the investigation of the molecular forces in gases using various phenomena*" [1, Doc. 102].

### 2.3. *First ideas on the kinetic nature of light in April-May 1901 compatible with Planck's quanta and Poincaré's paper in Lorentz' Festschrift?*

In 1993, Renn [3] suggested the original point of view according to which Einstein's ideas about light quanta could be traced back to 1901, on the basis of his atomistic vision (in relation to the molecular forces theory, capillarity, dissociation, electrical conduction in metals, etc.), his memories, and his letters to Mileva. Indeed, during his discussions with Shankland in 1952, Einstein said "*that the photoelectric effect paper was also the result of five years pondering and attempts to explain Planck's quantum* [that he learned in April 1901 [1, Doc. 97]] *in more specific terms*" [8]. The letter addressed to Besso in 1951 also confirms the 1901 year: "*All these fifty years of persistent speculations did not approach me to the answer "what light quanta are"* "[6, L. 177]. More precisely, by the end of April 1901, Einstein told Mileva about his intuition « *that when light is generated, direct conversion of motional energy to light may take place because of the parallelism kinetic energy of the molecules – absolute temperature – spectrum (radiating space energy in the state of equilibrium)* » [1, Doc. 102], bringing into play Wien's law of 1896 and the *ad hoc* introduction of proportionality between the kinetic energy of molecules and the radiation frequency. If Einstein considered that the kinetic origin of the emission was to be found in the radiation in itself, we can easily understand his enthusiasm when he read Lénard's article, an enthusiasm that he shared with Mileva in 1901: « *I have just read a marvellous paper by Lenard on the production of cathode rays by ultraviolet light. Under the influence of this beautiful piece of work, I am filled with such happiness and such joy* […] » [1, Doc. 111]. Indeed, in this absorption process, kinetic light energy could be returned to electrons.

It should be noted that Michele Besso reminded Einstein in 1945 that their first discussion, fifty years before, was about the nature of light "*between Newton and Huygens*" [6, L. 145]. If Einstein had a first idea about light quanta in 1901, he probably also pictured them, being a physicist, as something that could only be a wave packet, or, as he later said, a "*light complex*" [9]. This naïve empirical model of light quanta allowed him to translate the property *E=hν* that Planck discovered for his resonators, as a property of the radiation itself. It should be noted that, in quantitative terms, Einstein knew

Stefan's law $E = Cte \times VT^{\,4}$ and Wien's law $\lambda T = Cte$, and that it is sufficient to assimilate a light quanta to a wave train of volume $V = Cte \times \lambda^3$ to obtain the proportionality relation of energy $E$ to frequency $\nu$ (*cf.* note 88 in [5]). Note that, at the end of his 1904 paper [10], when Einstein obtains Wien's relation by identifying the mean energy to its fluctuations for the black body radiation, he argues that it applies to a cavity of size of order $\lambda$. It is likely that this kinetic vision of energy proportionality to frequency could have been reinforced by the reading of Poincaré's article entitled *Lorentz theory and the principle of Reaction* [11], also published in the *Festschrift*, an article that Einstein could have discovered at the same time as Planck's. Poincaré then introduced the momentum density of radiation and applied it to a Hertzian oscillator placed at the focus of a parabolic mirror. Not only did he establish the law of transformation of the energy of a light impulse, with a change of the inertial frame of reference, using the Lorentz transformations of 1895 (at first order of $V/c$) as Miller noticed [12], but also the transformation of its length (and thus also of its wavelength). The relations given by Poincaré thus imply that the frequency and energy of a plane monochromatic wave train transform in the same way. It is not unlikely that it is the reading of the *Festschrift* at the Lombardo Institute library, that could have lead Einstein to write to Mileva on April 14[th] that his « *views on the nature of radiation have again sunk back into the sea of haziness"* and lead him, with the parallel and simultaneous reading of Planck's 1901 article introducing energy quanta [13], to a dual vision of light.